# Universal Direct Proportionality:

# A New Analysis of Quantum Hall Effect

Timir Datta[1], Ming Yin[2] and Michael Bleiweiss[3]

1) University of South Carolina, Columbia, SC 29208.
2) Benedict College, Columbia, SC 29204.
3) Naval Academy Preparatory School, Newport, RI 02841.

**ABSTRACT**

This article addresses the researchers studying the magneto-transport properties of unconventional systems such as bulk samples and new materials or those working in the soft quantum limit. We have developed a procedure for the analysis of data, so as to increase the confidence level about the presence of quantum hall effect in these results. It is shown that for both integer and fractional QHE's, irrespective of temperature, sample material and sign of the charge carriers, a universal direct proportionality exists in the dimensionless variables $B^*_{min}$ and $R^*_{xy}$ (defined in the text). This procedure does not pre-assign the values of the Landau-level filling factors ($\nu$); but determines both ($\nu$) and carrier density ($n_s$) in a statistically significant manner. We also find a natural definition of the regime of the soft -limit. Results of the analyses of several sets of literature and synthetic data are reported.

Ever since the discovery of integer and fractional quantum hall effects (QHE) [1-4] our understanding of "matter in the extreme quantum limit" or $\beta \gg 1$ has improved at a phenomenal rate. The parameter $\beta$ is equal to the ratio of the magnetic and thermal energies ($B\mu_B/k_BT$); here $\mu_B$ is the appropriate effective Bohr magneton and $k_B$ is the Boltzman constant, temperature is T and B is the applied magnetic field. Currently there has been a resurgence of interest[5] in studying the behavior of quantum hall states in the soft limit that is when $\beta$ is small. Furthermore, explorations for QHE also continue at an ever-increasing pace into new non-heterostructure materials such as bulk matter[6,7], replica opals[8] and other systems[9,10].

In figure 1 is a von Klitzing plot of a set of typical for the hall resistance $R_{xy}$ and magneto resistance $R_{xx}$ isotherms as functions of B. Traditionally, in such a plot, each of the steps in the hall curve ($R_{xy}$) is marked by a Landau-level filling factor ($\nu$). In point of fact, often the filling factor has been used as the quantitative metric on the hall curve.

Analysis of the low field data in particular is the subject of this article. In figure 1 we notice that in the low field region the hall plateaus in $R_{xy}$ might not be well defined but vestiges appear to be present, this is supported by the clear Subnikov-deHass (SdHass) oscillations in the magneto resistance ($R_{xx}$) data. The difficulty is compounded by the fact that there may be no correlation between apparent steps in the hall curve with features in the SdHass curve such as at points B and C or the hall steps may not be sufficiently flat but sloped as at point "D". For a step on the hall curve to represent a quantum hall signal this step has to be associated with an SdHass feature. In actual situations with constant instrumental resolution the low field high temperature data, may show a higher proportion of noise and relative uncertainty than those at the higher values of β.

Deviations from the ideal flat plateaus and the presence of factional ν states may render the identity of ν, ambiguous. Because, when the generalized von Klitzing relationship is applied with both integer and fractional filling factor, ν:

$$R_{xy} = (\frac{h}{e^2})\nu^{-1} \qquad \ldots 1$$

Where h is Planck's constant and e the electron charge. It is possible that within the resolution of the data, from the experimentally measured values of $R_{xy}$ equation 1 may be satisfied by more than one, say two values of ν, namely $\nu_1$ and $\nu_2$. Example of such a situation is the rational fractions 7/5 and 10/7, both of these states are observed and are well known. As decimals one is 1.400 and the other is 1.428; so, within 2% they are numerically the same. However, in the nomenclature of quantum hall literature, ν = p/q is the pth-order Landau-sate in the qth-series. In addition as sample quality improves the hall curves show ever more details reveling fractal characteristics such as self-similarity[11]. It has been suggested that nested inside the 1/3 and 2/5 states there are a whole range of states with filling factors such as 5/13, 8/21 … 4/11 etc[12,13].

Confronted with such indeterminacy the confidence level of the data set is reduced and one may even question if the steps in the hall data are associated with quantum effects in the first place. However, perusing the totality of the hall and SdHass isotherms of figure 1, especially the correlations between them, it would appear highly unlikely that the mutually synchronized patterns are mere random coincidences and not evidence of QHE. The uncertainty is more pronounced if the available data is limited only over the low field regime, such as those contained in the small brown rectangle marked in the lower left hand corner of figure 1. Hence, the primary issue is how to be confident that the "wiggles" and steps in the experimental data are indeed evidence of quantum hall effect. A procedure that is free of the assignment filling factors and that would attest to the QHE character of the experimental results as a whole will be very useful.

Let us ask, what features in the von Klitzing plot are essential and must be retained in any proposed analysis? We intend to incorporate quantitative information from both the diagonal and off-diagonal transport coefficients; this is applied globally, i.e., not to individual points (steps) but to all the relevant QHE signatures. Clearly, the resistance steps on the hall curve are important but what about the SdHass curve? We reason, it is the period or magnetic field at the SdHass minima that contains the critical information. The procedure should satisfy a double consistency condition: (i) first, the hall resistance ($R_{xy}$) steps apply to equation 1. (ii) Secondly, the condition relating the charge carrier density ($n_s$) magnetic field at the SdHass minima ($B_{min}$) and inverse $\nu$ is maintained, that is:

$$B_{min} = (\frac{n_s h}{e})\nu^{-1} \qquad \ldots 2$$

The two simultaneous linear dependences of both $R_{xy}$ and $B_{min}$ as determined by equations 1 and 2 may be combined into a (single) direct proportionality between the $B_{min}$ and $R_{xy}$. Because from equations 1 and 2 it follows that,

$$B_{min} = (en_s)R_{xy} \qquad \ldots 3a$$

Or equivalently,

$$R_{xy} = (en_s)^{-1} B_{min} \qquad \ldots 3b$$

The proportionality constant $(en_s)$ is the aerial charge density, $\rho_e$, of the QH fluid. Numerically, $\rho_e$ is equal to the value of the field, $B_{min}(1)$, when $R_{xy} = h/e^2$ or the field for the unit filling factor. An additional advantage of equations 3a and 3b is that they contain only the directly measured quantities namely $B_{min}$ and $R_{xy}$ and as promised do not involve $\nu$. Once the carrier density $n_s$ is calculated the values of the filling factor ($\nu$) for each of the hall plateaus can be determined by inverting equation 2, that is

$$\nu = (\frac{en_s h}{e^2})(B_{min})^{-1} \qquad \ldots 4$$

or

$$\nu = 25812.8 * (en_s)(B_{min})^{-1} \qquad \ldots 5$$

Although the $\nu$'s obtained from equation 5 are not necessarily unambiguous; all the $\nu$'s are determined within the confidence level consistent with the entirety of the experiment.

In table-1 we show some of the quantitative data from figure 1. The resistance values entered in red are for those steps that (within the depicted experimental resolution) are "perfectly flat" or represent the constant hall plateaus; the blue data are from the low β end. The value of $B_{min}$ would depend on its operational definition, such as the mid point field value, between the two SdHass zeros (B- and B+ in figure 1) or that between the two half resistance points etc. Our data (table-1) contains about 2-5% error; an actual experiment can have less errors for high β and higher errors in the low β regime. In figure 2, the data of table-1 are plotted as per equation 3a, that is $B_{min}$ as a direct function of the hall resistance. The red data points were weighted three times more than the others in determining the straight line fit and the slope represents a charge density of 8.03 x$10^{-4}$ C/$m^2$, which corresponds to about 5x$10^{15}$electrons/$m^2$. The calculated values of ν corresponding to this slope are shown on the last column of table-1.

Equations 3a and 3b is key in all quantum hall effect experiments, they require that with every thing else held constant to observe the hall steps at low fields one would need low charge or carrier density. It is interesting to relate our result to the parameter $β = Bμ_B/k_BT$, i.e, $B_{min} = β_{min}(k_BT/μ_B)$ hence from equation 3a we obtain:

$$β_{min} = (en_s)(\frac{μ_B}{Tk_B})R_{xy} \qquad …6a$$

The equation above contains all the relevant parameters of the problem, the two relating to the material $n_s$ and $μ_B$ and the experimental quantity the temperature T. From equation 6a we may define $β_{min}(1)$, the reduced field at SdHass minimum associated with the unit quantum hall step, i.e.,

$$\frac{Tβ_{min}(1)}{μ_B n_s} = (\frac{h}{ek_B}) \qquad …6b$$

Expressing the effective Bohr magneton of the charge carriers ($\mu_B = e\hbar/2m^*c$) in terms of their effective mass $m^*$ we may rewrite equation 6b as

$$T\beta_{min}(1)(\frac{m^*}{n_s}) = (\frac{h^2}{4\pi c k_B}) \qquad \ldots 6c$$

Equations 6a-6c show that when $\beta$ is small every thing else being equal, the condition of the observation of QHE is that the effective $\mu_B$ for the system be also small or effective mass is large. Or in other words amongst two systems at the same temperature, in the system with small carrier density and large effective mass, QHE will be observable at a lower $\beta$.

Also, dividing the right hand side of equation 3a by the quantum unit of resistance ($h/e^2$) and the left hand side by the value of the field $B_{min}(1)$ defined earlier to be the B at the SdHass minimum associated with the unit hall resistance step (feature) then in these normalized units of field and resistance, B* and R*, we obtain:

$$\frac{B_{min}}{B_{min}(1)} = \frac{R_{xy}(\Omega)}{(h/e^2)}$$

or

$$B^*_{min} = R^*_{xy} \qquad \ldots 7$$

Equation 7 predicts that any set of QHE data when plotted in these reduced variables will follow a direct universal direct proportionality with unit slope, irrespective of the (positive or negative) sign of the carrier or their density. To test this behavior we digitized the QHE data from three classic papers by von Klitzing[14], Stormer[15] and Mendez[16] et al. Of the three papers the first[14] was for electronic integer effects, the second[15] was also electronic but for fractional effects and third[16] was hole conduction for fractional effect. Table 2 shows some of the data from these sources. Note that we obtained the values of each of the individual points from published figures in the references, so there were several sources of digitization errors. Graphical analysis of these data reveled that the carrier density, $n_s$ for the first set was $(3.4± 0.4) \times 10^{15}$ electrons /m$^2$, that for the second set $(2.9± 0.3) \times 10^{15}$ electrons /m$^2$. The published[16] value for the third set is $3.07 \times 10^{15}$ holes /m$^2$ in comparison we estimated it to be $(3.03± 0.1) \times 10^{15}$ holes /m$^2$. The resulting plot from all the three sources is shown in figure 3.

Clearly the results from three different quantum hall situations namely, the low field or high integer QHE behavior at a temperature of 8mK [14]; the wide low and high fractional states from ref (15) and the hole data taken at 0.51K[16] all follow the direct proportionality predicted by equation 7. Coincidentally the filling factor of the last red circle was not identified in reference15, however with the help of equation 5 we assign it a value of ν equal to 5 (± 0.1).

In terms of its parameter space, figure 3 also provides a new operational definition of the quantum soft limit, namely in these dimensionless units the regime of $0>B^*>>1$ and $0>R^*>>1$ The behavior in this soft limit are shown in figure 4, here the data of table-1 are included unfortunately the data from set two did not fall in this regime and cannot be seen. Here data from table-1 are indicated by the red circles, those from the set one by green diamonds and set three are shown as triangles. The directly proportional scaling persists even in the soft limit region. The closeness of the value of the slope to unity may be seen from the 1:1 ratios in x:y coordinates of the data points. However a least square fit increases the reliability of each point and the accuracy of the digitization process. Such a statistical procedure enhances the confidence level of the entire data set well beyond the precision of individual measurements.

In this article our goal has been to develop a procedure for the analysis of data obtained in magneto-transport experiments, so as to increase the confidence level about the presence of quantum hall effect. This is a problem faced not only by researchers in the soft-limit but also by those who are searching for QHE in unconventional systems such as bulk samples and new materials. We describe that even when the hall curve shows no "really flat" steps and the values of filling factors are ambiguous, by correlating the magnetic fields (B) at the SdHass minima with the hall resistance steps at it is possible to improve the confidence level of the QHE signatures. It is also shown that irrespective of temperature, sample material and sign of the charge carriers, a universal direct proportionality exists in dimensionless variables $B^*_{min}$ and $R^*_{xy}$ for both integer and fractional quantum hall effects. We find a natural, operational definition of the soft quantum limit. Pre-assignment of filling factor values (ν) are not required instead estimates of ν, the surface charge density and the electron (hole) concentration in the quantum hall condensate are obtained in a statistically significant manner.

## Acknowledgements

This work was initiated by a DARPA subcontract from Honeywell Corporation, and is currently supported in part by the USC, Nanocenter.

**Table -1**

| $R_{hall}$ (kOhm) | Nominal $\nu$ | Bmin(Tesla) | Calc $\nu$ |
|---|---|---|---|
| **8.72** | 3.00 | 7.00 | 2.99 |
| **6.45** | 4.00 | 5.21 | 3.97 |
| **5.21** | 5.00 | 4.18 | 4.97 |
| 4.42 | 5.85 | 3.54 | 5.86 |
| **3.74** | 7.00 | 2.96 | 6.99 |
| 3.28 | 7.86 | 2.64 | 7.84 |
| **2.66** | 10.0 | 2.11 | 9.86 |
| 2.26 | 11.4 | 1.79 | 11.6 |
| 1.93 | 13.4 | 1.54 | 13.5 |
| **1.64** | 15.7 | 1.32 | 15.7 |
| **1.47** | 17.5 | 1.18 | 17.6 |
| **1.30** | 19.8 | 1.04 | 20.0 |
| **1.08** | 24.0 | 0.89 | 23.2 |

## Table-2

| Source | Rhall (h/e$^2$) | Bmin (tesla) | B*min |
|---|---|---|---|
| Ref. 14 | 0.33134 | 5.2998 | 0.33459 |
| " | 0.24936 | 3.8660 | 0.24406 |
| " | 0.16420 | 2.5794 | 0.16284 |
| " | 0.12321 | 1.9381 | 0.12235 |
| " | 0.095858 | 1.5258 | 0.096324 |
| " | 0.079936 | 1.3196 | 0.083309 |
| " | 0.069403 | 1.1300 | 0.071337 |
| " | 0.055441 | 0.95786 | 0.060471 |
|  |  |  |  |
| Ref.15 | 2.9203 | 28.540 | 3.0153 |
| " | 2.4526 | 23.930 | 2.5283 |
| " | 2.2881 | 22.193 | 2.3447 |
| " | 2.1781 | 21.370 | 2.2578 |
| " | 1.7705 | 17.060 | 1.8024 |
| " | 1.7119 | 16.600 | 1.7538 |
| " | 1.6469 | 15.765 | 1.6656 |
| " | 1.4783 | 14.125 | 1.4923 |
| " | 1.2417 | 11.970 | 1.2647 |
| " | 1.0000 | 9.4650 | 1.0000 |
| " | 0.76692 | 7.0800 | 0.74802 |
| " | 0.61505 | 5.6000 | 0.59165 |
| " | 0.52322 | 4.7610 | 0.50301 |
| " | 0.35066 | 3.1300 | 0.33069 |
| " | 0.28507 | 2.3700 | 0.25040 |
| " | 0.21847 | 1.9255 | 0.20343 |
|  |  |  |  |
| Ref. 16 | 1.6933 | 21.442 | 1.6437 |
| " | 1.4976 | 19.273 | 1.4774 |
| " | 0.99999 | 13.045 | 1.0000 |
| " | 0.49952 | 6.4134 | 0.49164 |
| " | 0.34571 | 4.3030 | 0.32986 |
| " | 0.26476 | 3.1515 | 0.24159 |

**Figures**

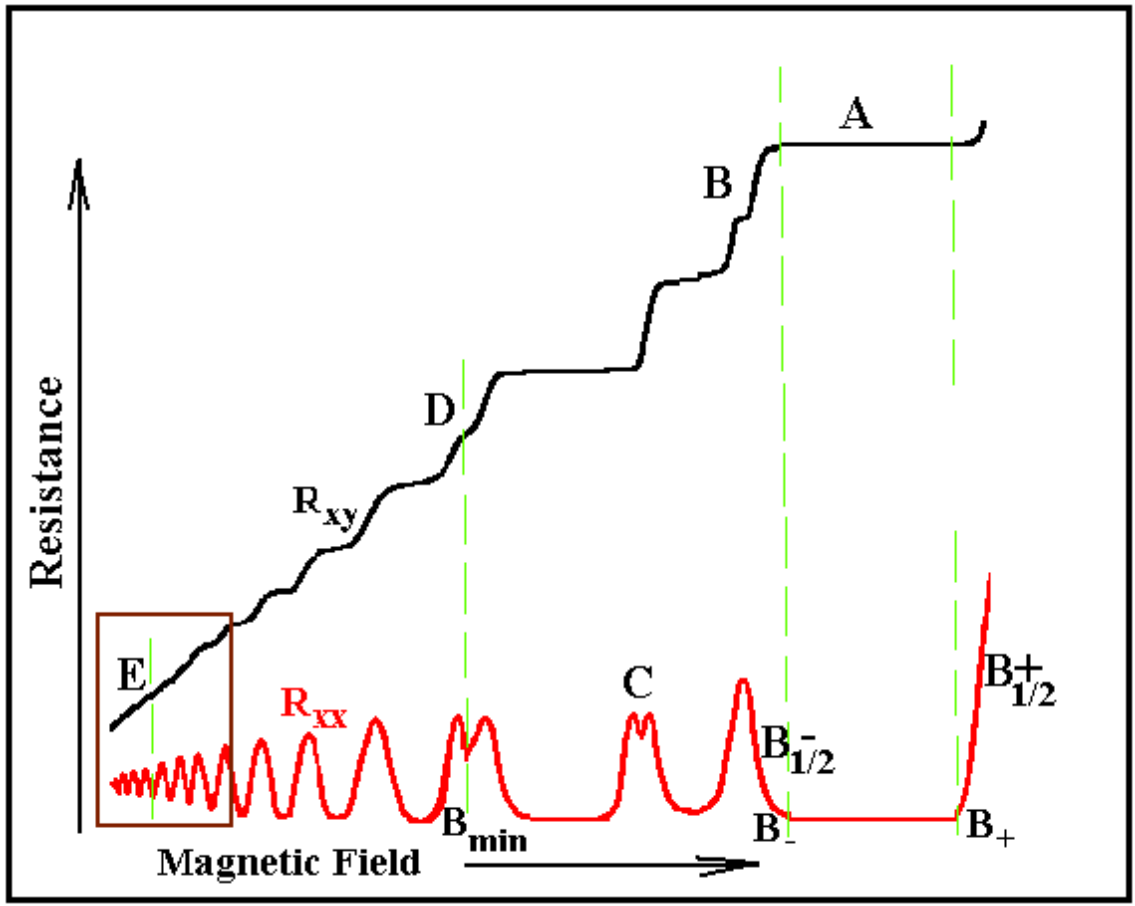

**Figure-1:** A "von Klitzing" plot from a hypothetical quantum hall experiment. The blue curve represents the hall ($R_{xy}$) data and the red curve that of magneto resistance ($R_{xx}$). The filling factor ν associated with the "perfectly" flat Hall plateaus such as the one in the region marked "A" are readily identifiable. However those at points marked by the green lines at "D" and "E" might not be unique. Point "B" has a step in the hall curve but no feature is resolved in $R_{xx}$ and the opposite is true at "C". Both B and C will not contribute to our analysis.

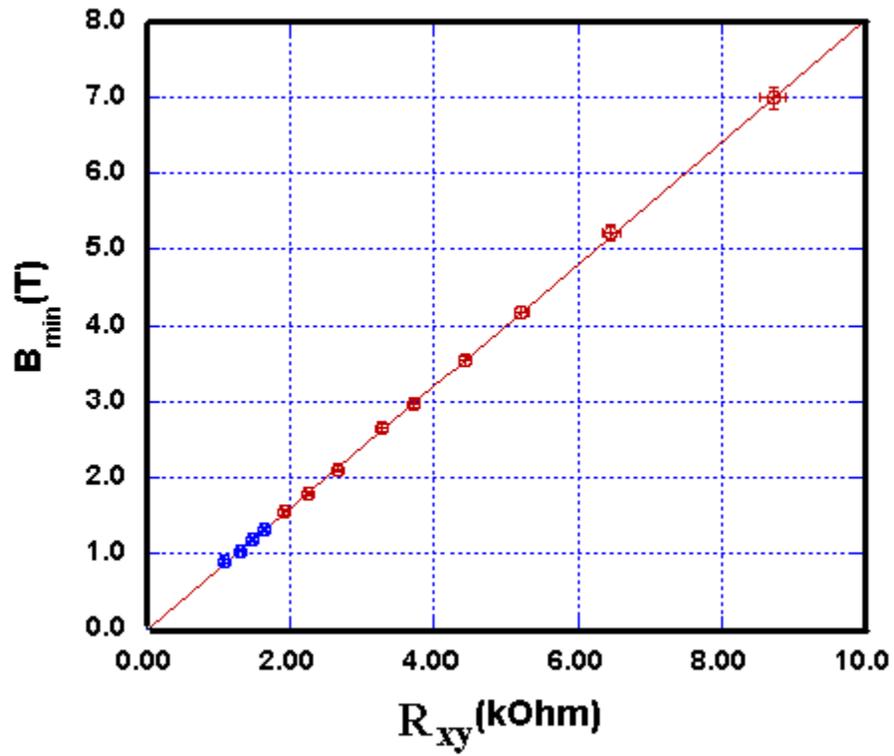

**Figure-2:** A $B_{min}$ vs $R_{hall}$ graph of the data of table 1. Clearly a direct behavior predicted by equation 7 is evident. Also the quality of the fit and hence the reliability of the last four (low β regime) points seen at the bottom left corner is quite satisfactory.

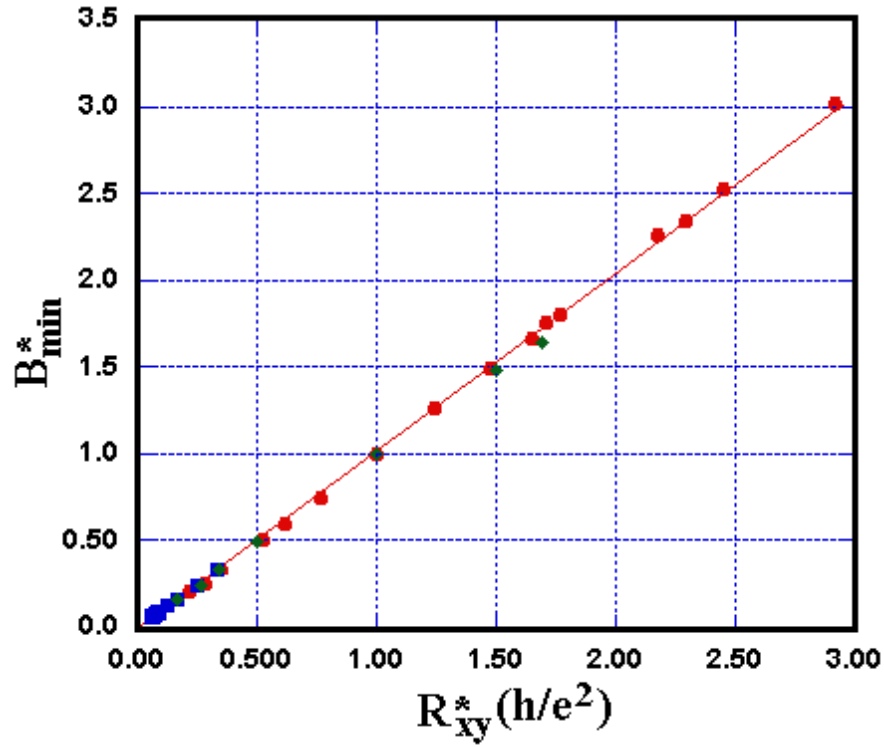

**Figure-3:** Universal direct proportionality (equation 7) in three different sets of published literature results listed in table 2. The blue squares are (8mK) from ref. 14, the red circles are from ref. 15 and the green diamonds (0.51K) are from ref. 16. The straight line is of unit slope with a correlation coefficient of 0.999 an excellent fit in view of the uncertainty in digitizing from figures in published sources.

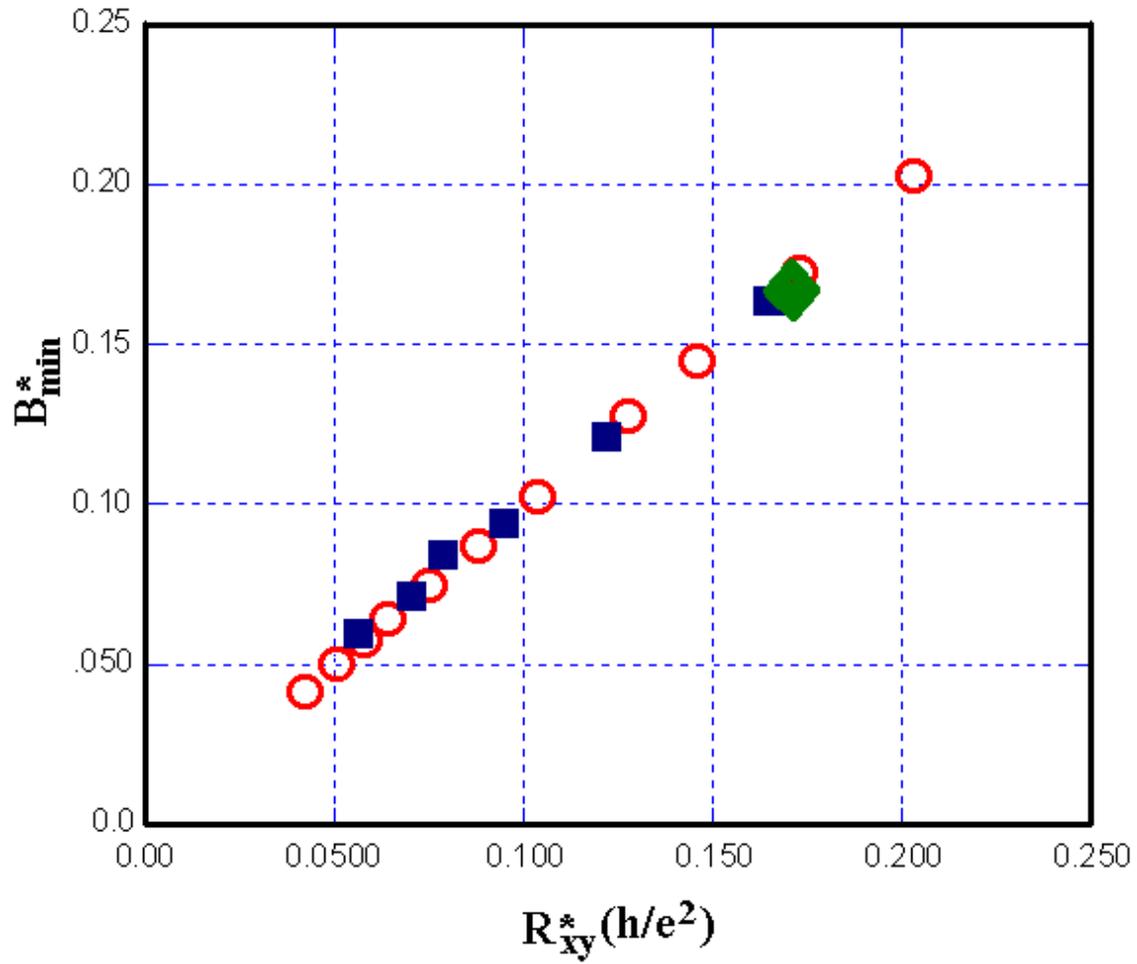

**Figure-4:** A magnified view of the directly proportional scaling behavior in the low β or soft limit region. Here data from table-1 (red circles) and the set one[14] (green diamond) and set three[16] (blue square) from table 2 are shown. The points from reference 15 are excluded because they correspond to a higher β regime.